\documentclass[twocolumn,showpacs,aps]{revtex4}

\usepackage{graphicx}
\usepackage{dcolumn}
\usepackage{amsmath}

\begin{document}


\title{Self-organized Criticality and Scale-free Properties in Emergent Functional Neural Networks}

\author{Chang-Woo Shin}
\email{shine@postech.ac.kr}
\author{Seunghwan Kim}
\email{swan@postech.ac.kr}
\affiliation{
Department of Physics, Pohang University of Science and Technology, and
Asia Pacific Center for Theoretical Physics,
San 31, Hyoja-dong, Nam-gu, Pohang, Gyungbuk, Korea, 790-784}

\date{\today}

\begin{abstract}
Recent studies on the complex systems have shown that
the synchronization of oscillators including neuronal ones is faster, stronger, and more efficient in the small-world networks
than in the regular or the random networks,
and many studies are based on the assumption that the brain may utilize the small-world and scale-free network structure.
We show that the functional structures in the brain are self-organized to both the small-world and the scale-free networks by
synaptic re-organization by the spike timing dependent synaptic plasticity (STDP),
which is hardly achieved with conventional Hebbian learning rules.
We show that the balance between the excitatory and the inhibitory synaptic inputs
is critical in the formation of the functional structure,
which is found to lie in a self-organized critical state.
\end{abstract}


\maketitle

\begin{figure*}[ptb]
\unitlength1cm
\resizebox{4.5cm}{4.2cm}{\includegraphics{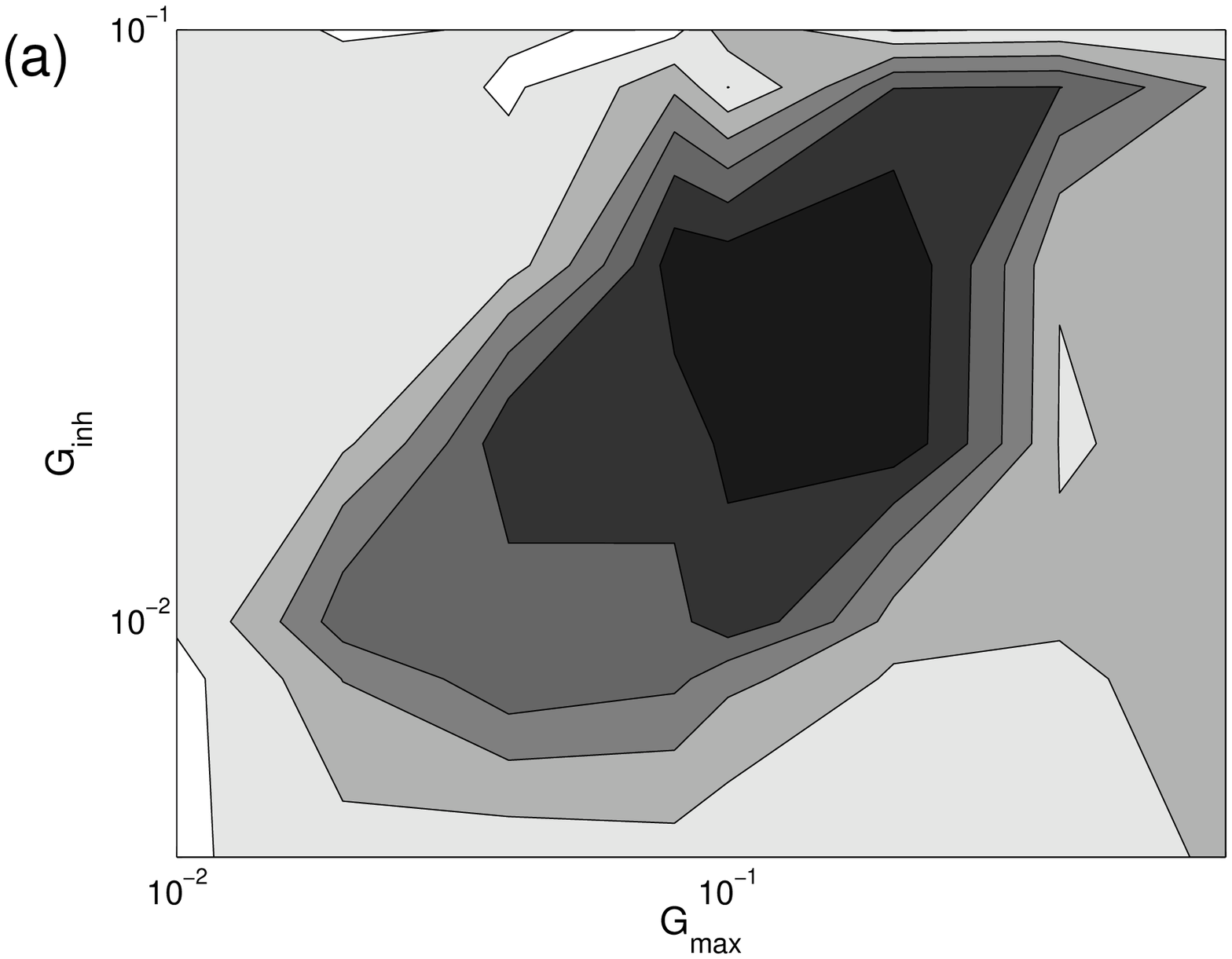}} ~~~~~
\resizebox{4.5cm}{4.2cm}{\includegraphics{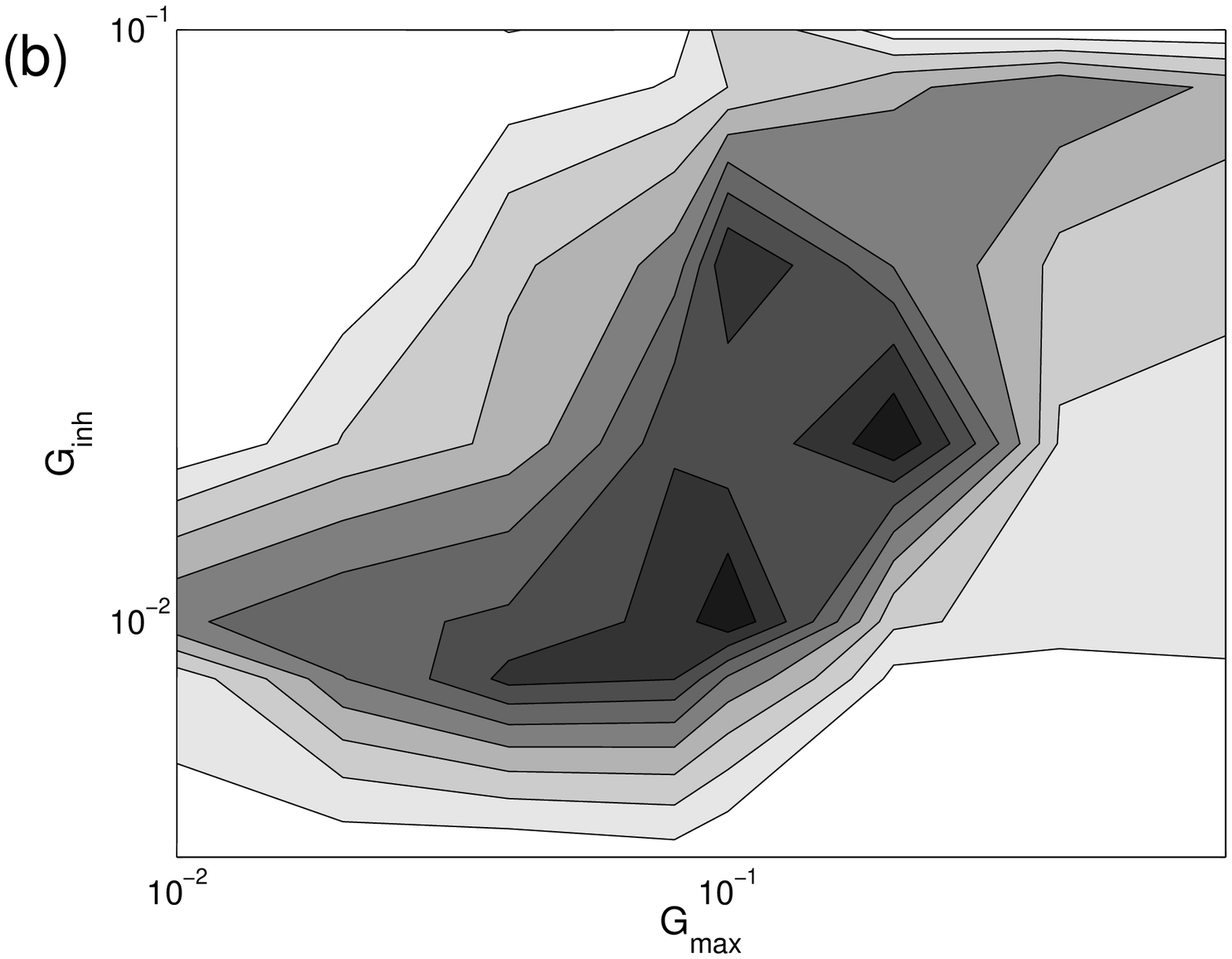}} ~~~~~
\resizebox{4.5cm}{4.2cm}{\includegraphics{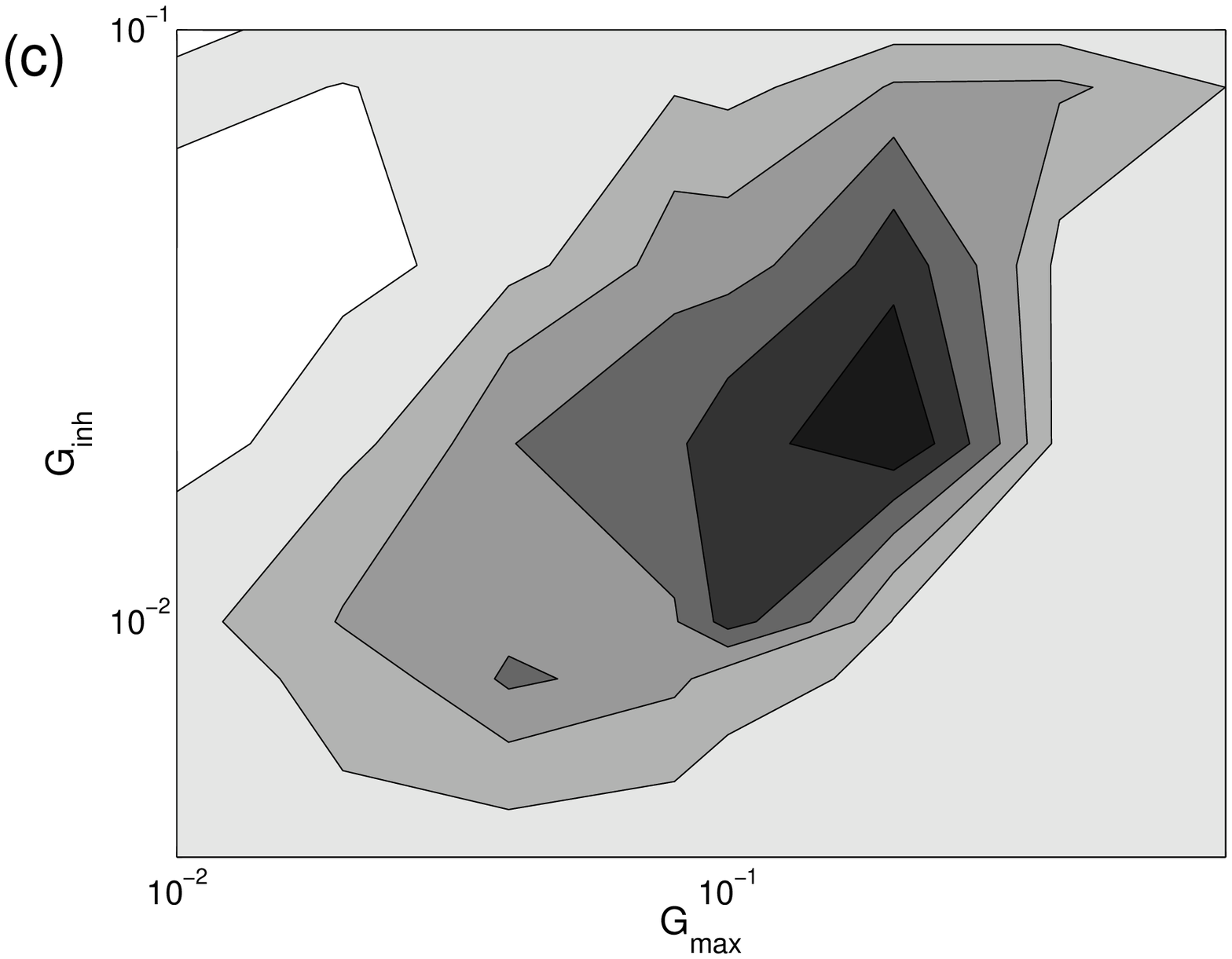}}
\caption{(a) Connection probability, $\left< k / N \right>$,
in the range between 0.03 (dark) and 0.6 (light),
(b) phase coherence, $\Gamma$, between 0.4 (dark) and 1.0 (light), and
(c) the ratio of the clustering coefficient to the case of the random network with the same connection probability, $C / C_{rand}$,
between 0.5 (light) and 7.5 (dark)
in the parameter space of $G_{max}$ and $G_{inh}$.
The results are averaged over 10 different initial conditions.
}
\label{FIG_DIAGRAM}
\end{figure*}

The brain is one of the most challenging complex systems.
The neurons, massively inter-connected to each other, show highly complex and correlated responses to the external stimuli,
which help the brain to extract relevant patterns from sensory inputs,
coordinate movements and control behaviors.
To understand the complexity of the nervous system,
we need to characterize its network structure
on which the spatio-temporal firing activities are supported.

Recent studies on the complex networks in a variety of systems describe real networks
by simply defining a set of nodes and connections between them.
Examples range over social networks, information networks, technological networks, and biological networks\cite{NEWMAN:2003:SIAMREVIEW}.
They lie between regular networks and fully random networks.
A wide variety of such systems are scale-free,
where the connectivity distributions take a power-law form,
and the topology and evolution of such networks
are governed by the common mechanism
such as the preferential attachment and the growth
regardless of the detailed nature of the specific networks \cite{NEWMAN:2003:SIAMREVIEW,ALBERT:2002:RMP,DOROGOVTSEV:2002:ADVPHY,AMARAL:2004:EURPHYSJB}

In the context of the complex network,
the topological structure of simple nervous systems, for example,
in the worm {\it Caenorhabditis elegans} neural network\cite{WHITE:1986:},
have been proved to be an inhomogeneous small-world network.
However, for the network in the brain,
more important is the functional structure than the morphological one
because the former is the direct carrier of the neuronal information in the form of spikes.
Moreover, the functional structure changes through the adaptive variation in the synaptic conductances
due to the inputs from external stimuli and
the internal dynamics of neurons in the network,
which in turn leads to the change in the responses of the network.
This feedback process of synaptic modification in the brain is believed to be
closely connected to the learning and memory.
Recently, this kind of synaptic changes have been observed
experimentally in various brain regions,
such as neocortical slices\cite{MARKRAM:1997:SCIENCE},
hippocampal slices\cite{DEBANNE:1998:JPHYSIOL} and cell cultures\cite{BI:1998:JNEUROSCI},
and the EEL of the electric fish\cite{BELL:1997:NATURE},
where long term synaptic modifications, both long term potentiation (LTP) and long term depression (LTD),
arise from repeated pairings of pre- and post-synaptic action potentials.
The sign and the degree of synaptic modification depend on their relative timing,
called the {\em spike timing dependent plasticity} (STDP).
For example, in the hippocampal CA3 region and neocortical slices,
by STDP a synapse is strengthened
if the presynaptic spike is followed by postsynaptic action potentials
within about $50ms$
and weakened if the presynaptic action potential follows postsynaptic spikes.

In this Letter,
we report that STDP reorganizes a globally connected neural network
spontaneously into a functional network
which is both {\it small-world} network and {\it scale-free}.
This complex network arises when the excitatory and inhibitory connection strengths
between neurons are balanced.
The neuronal activities on this small-world scale-free neural functional network is found to lie
in a self-organized critical state.
The small-world scale-free functional structure is formed for a wide class of neuron models
including the Hodgkin-Huxley (HH) model, which we have also tested
and in a wide range of control parameters, such as the strength of the external stimulus, and parameters related to STDP,
independent of the initial conditions.
The neuronal oscillators in the functional structure with a small connection probability
organized by STDP show fast synchronous responses to the external stimuli,
which implies that
STDP give the neural network both the reliability in information transformation and the stability preventing from epileptic over-excitation.
It is noticeable that
the functional structure is formed depending on the spatio-temporal dynamics of the neurons
rather than explicit preferential attachment rule.

As a model neuron, we take the FitzHugh-Nagumo (FHN) model \cite{FITZHUGH:1961:BJ},
which is a two dimensional relaxation oscillator with two time scales
but contains the essential ingredients of nervous excitation
and fast action potential generation followed by a slow refractory period:
\begin{eqnarray}
\epsilon \dot{v} &=& I_{ion} + I_{syn} + I_{ext} \nonumber \\
\dot{w} &=& v - w - b \label{EQ_FHN} \\
I_{ion} &=& v (v-a) (1-v) - w ,\nonumber
\end{eqnarray}
where, with $\epsilon \ll 1$, $v$ is a fast voltage-like variable,
$w$ a slow recovery variable,
$I_{ion}$ the ionic current through the membrane with cubic nonlinearity,
$I_{ext}$ the external current stimulus.
The synaptic current input to the $i$-th neuron is
the sum of excitatory and inhibitory currents from pre-synaptic neurons:
\begin{equation}
I_{syn}(t) = \sum_{j \ne i} \left[ g_{ij}(t) (V - v_{i}(t)) + \bar{g}_{ij}(t) (\bar{V} - v_{i}(t)) \right] ,
\end{equation}
where $g_{ij}$ ($\bar{g}_{ij}$) is the excitatory (inhibitory) synaptic conductance from the $j$-th neuron to the $i$-th neuron
and $V$ ($\bar{V}$) the excitatory (inhibitory) synaptic reversal potential respectively.
If $j$-th pre-synaptic neuron makes an action potential at time $t^{*}$, it increases the post-synaptic conductances
by the amount of the coupling strength of the synapse at $t^{*}$ normalized by the number of neurons,
$g_{ij} \rightarrow g_{ij} + G_{ij}(t^{*}) / (N-1)$ and $\bar{g}_{ij} \rightarrow \bar{g}_{ij} + \bar{G}_{ij}(t^{*}) / (N-1)$,
and the synaptic conductances decay exponentially:
\begin{equation}
\tau_{syn} \frac{d{g}_{ij}}{dt} = - g_{ij} ~ \mbox{ and } ~
\bar{\tau}_{syn} \frac{d \bar{g}_{ij}}{dt} = - \bar{g}_{ij} .
\end{equation}

In our STDP neural network model, we assume that inhibitory synaptic coupling strengths remain constant \cite{BI:1998:JNEUROSCI}, $\bar{G}_{ij}(t) = G_{inh}$,
while excitatory synaptic strengths change multiplicatively at every firing events \cite{Rossum:2000:JNEUROSCI,RUBIN:2001:PRL,BI:2001:ANNUREVNEUROSCI}:
\begin{equation}
\Delta G_{ij} = G_{ij} \cdot W (\Delta t) .
\end{equation}
The amount of the synaptic modification by STDP
depending on the time difference between pre- and post-synaptic spikes, $\Delta t = t_{post} - t_{pre}$,
is modeled by the STDP modification function:
\begin{equation}
W(\Delta t) = \left\{
\begin{array}{c}
A_{+} \exp \left( - \Delta t / \tau_{+} \right) ~~ \mbox{if} ~ \Delta t > 0 \\
- A_{-} \exp \left( \Delta t / \tau_{-} \right) ~~ \mbox{if} ~ \Delta t < 0 
\end{array}
\right.
\end{equation}
and $W(\Delta t = 0) = 0$.
The parameters $\tau_{\pm}$ determine the temporal window of the spike intervals,
and $A_{\pm}$ determine the maximum amount of synaptic modification.
It has been shown experimentally that in most situations, $A_{+} > A_{-}$, $\tau_{+} < \tau_{-}$,
and the integral of the function $W$ is usually negative\cite{BI:2001:ANNUREVNEUROSCI}.
Here, the parameter values
are chosen to be $A_{+}=0.01$, $A_{-}=0.006$, $\tau_{+}=1.0$, and $\tau_{-}=2.0$.
$G_{ij}$ lies in $0 < G_{ij} \leq G_{max}$
and if $G_{ij}$ increases over the maximal value,
$G_{ij}$ is set to $G_{max}$.
Other parameters are set to
$a = 0.5$, $b = 0.12$, $\epsilon = 0.005$,
$V = 0.7$, $\bar{V} = 0.0$, and $\tau_{syn} = \bar{\tau}_{syn} = 0.2$.


\begin{figure*}[btp]
\resizebox{4.2cm}{4.2cm}{\includegraphics{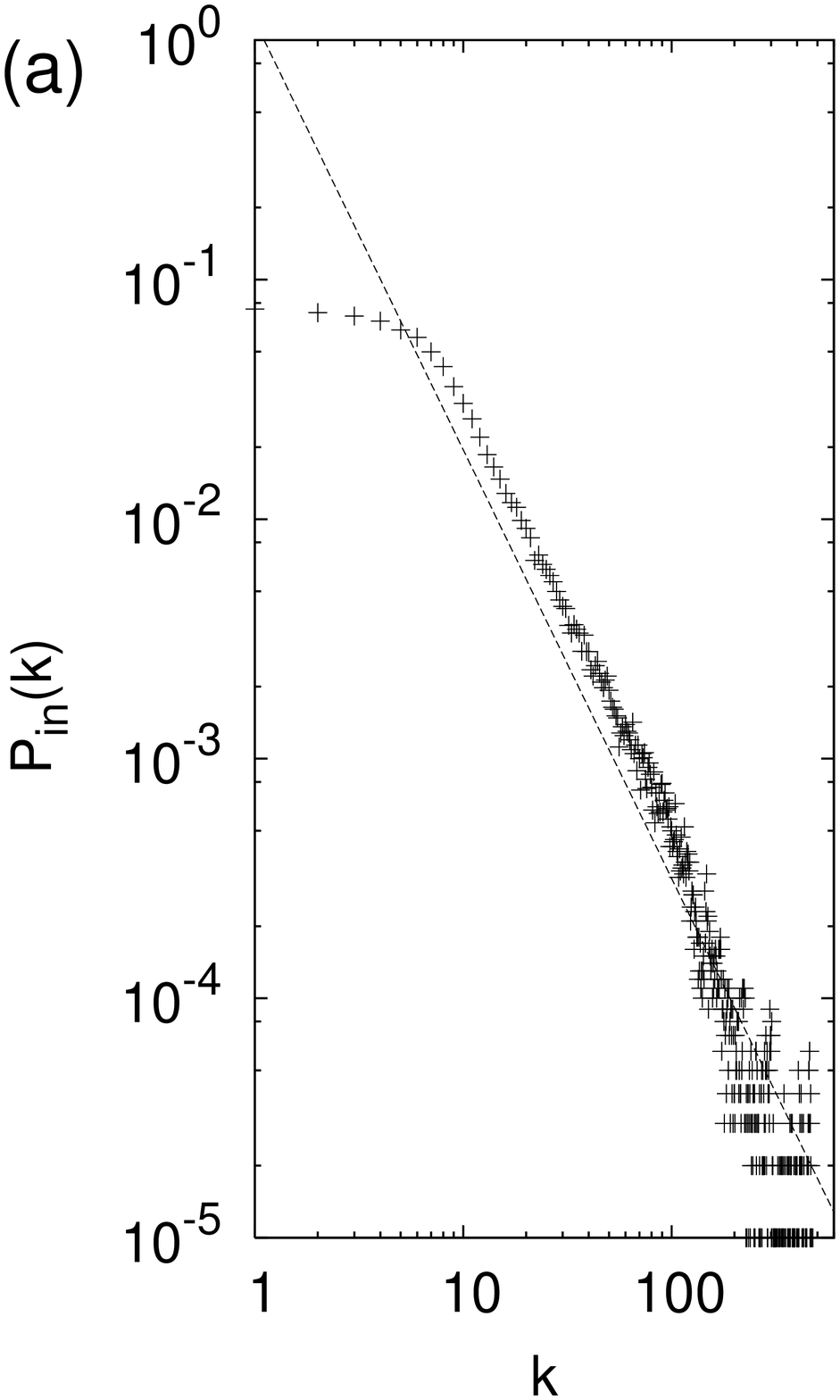}} ~~~~
\resizebox{4.2cm}{4.2cm}{\includegraphics{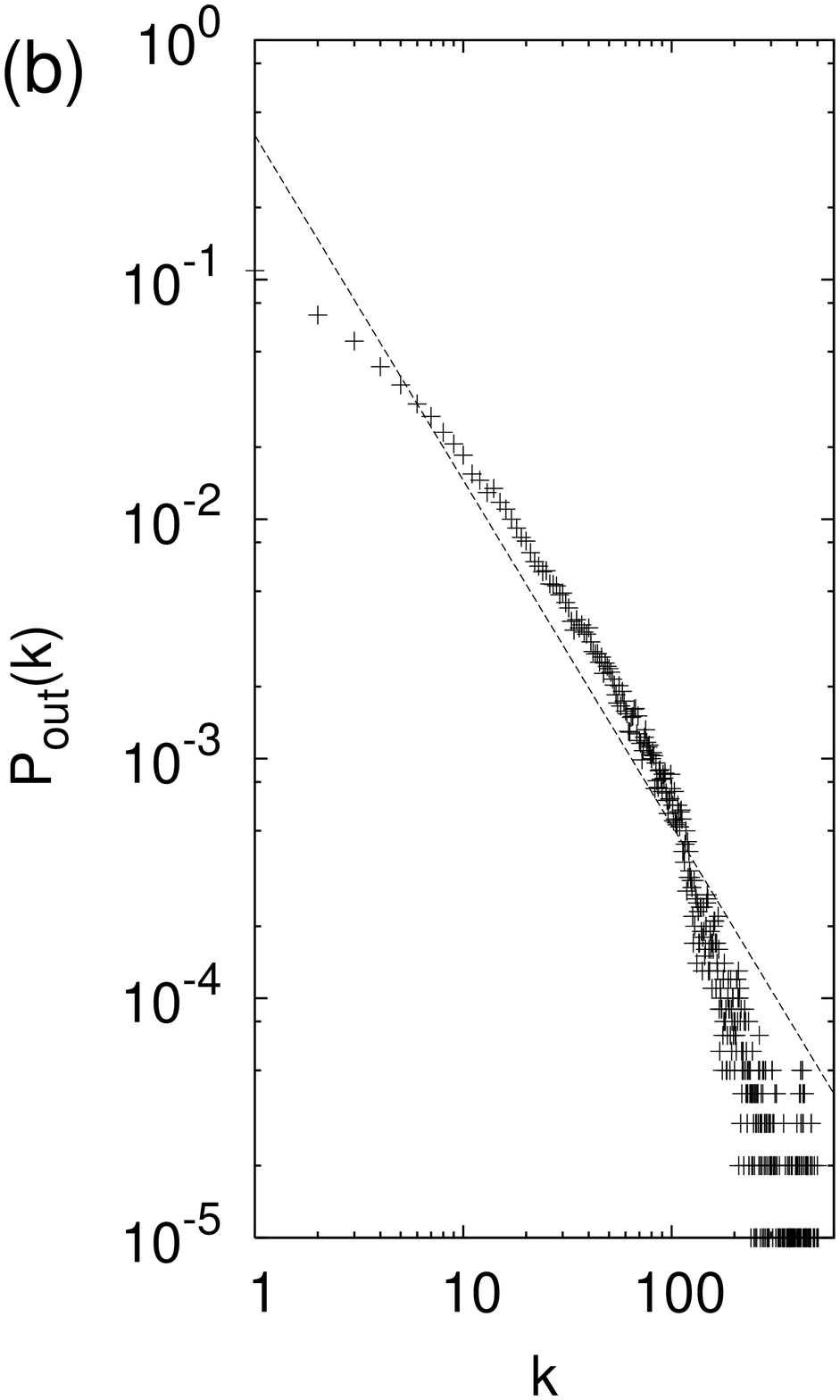}} ~~~~
\resizebox{4.2cm}{4.2cm}{\includegraphics{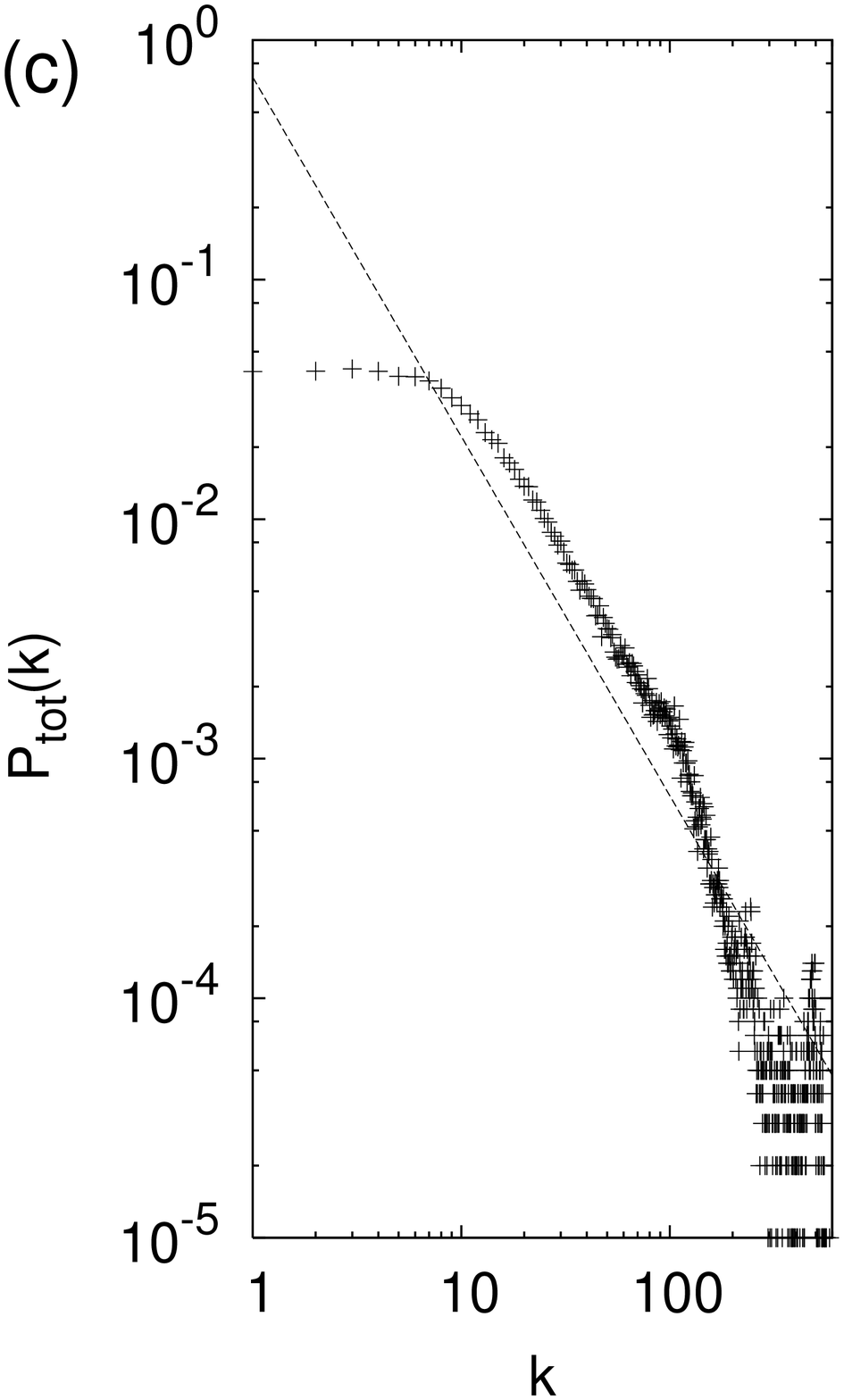}}
\caption{Log-log plot of degree distributions of the functional structure organized by STDP
when the excitatory and the inhibitory inputs are balanced.
In-degree probability distribution (a),
out-degree probability distribution (b), and
total-degree probability distribution independent of directions (c).
The scaling exponents are $\gamma_{in} \approx 1.7$, $\gamma_{out} \approx 1.4$, and $\gamma_{tot} \approx 1.5$
in the case of $G_{max} = 0.2$ and $G_{inh} = 0.02$.
The results are averaged over 100 runs with different initial conditions.
}
\label{FIG_DEGREE}
\end{figure*}

We start from the globally coupled network of $N=1,000$ neurons
with random initial coupling strengths of $G_{ij}$, $0 < G_{ij} \leq G_{max}$,
and investigate how the functional structure develops spontaneously in time.
Given the external $dc$-current, $I_{ext} = 0.2$, which is supra-threshold stimulus
for spontaneous generation of action potentials,
after a period of relaxation by STDP,
some population of synapses are strengthened to the maximum conductance value, $G_{max}$,
while the other population of synapses are weakened to near zero
and, therefore, become silent to their postsynaptic neurons.
This is similar to the bimodal distribution in the case of
the balanced excitation of synapses
from many input neurons to a single neuron \cite{SONG:2000:NATURENEUROSCI}.
As a result, even if the neurons in the network
are morphologically connected all-to-all by synapses,
the functional structure can be re-organized by STDP in which 
each neuron is functionally connected to only a small population of neurons.

Each synapse is regarded as functionally connected if the synaptic conductance
is larger than a critical value, $G_{ij} > G_{c}$,
and functionally not connected otherwise.
In Fig.\ref{FIG_DIAGRAM}(a),
the average connection probability,
the ratio of the number of strengthened synapses to the total number of synapses,
$\left< k/N \right>$, is shown
in the parameter space of $G_{max}$ and $G_{inh}$.
In this figure, there exists a region along the diagonal,
where the connection probability is very small,
and on either side of this region the connection probability becomes relatively large.
In accordance with the diagram,
three distinct classes of the network states can be identified:
synchronized, clustering, and dispersed network states.
To characterize the dynamical properties of the functional network organized by STDP,
we define the phase of a neuron at time $t$ between each firing time piece-wise linearly\cite{PIKOVSKY:1997:PHYSICAD}:
\begin{equation}
\phi(t) = 2\pi \left( \frac{t-t^{*}_{n}}{t^{*}_{n+1}-t^{*}_{n}} + n \right),
\label{INST_PHASE}
\end{equation}
where $t^{*}_{n}$ is the $n$-th firing time of the neuron.
The phase coherence, $\Gamma$,
of neurons in the network is defined as:
\begin{equation}
\Gamma = \max \left\{ \left| \lim_{T\to\infty} \frac{1}{T} \int^{T}_{0} \frac{1}{N} \sum^{N}_{j \ne k} e^{i n \Delta_{jk}(t)} dt \right| \right\}_n ,
\label{EQ_PHS_COH_MOD}
\end{equation}
where $\Delta_{jk}(t)$ is the difference of the instantaneous phases of $j$-th and $k$-th neurons at time $t$.
Note that $\Gamma$ saturates to $1$ if the firing times of all neurons are coherent with $n$-clusters and $0$ if they are random.
The dependence of the phase coherence, $\Gamma$, in Fig.\ref{FIG_DIAGRAM}(b)
is similar to the one for the average connection probability in Fig.\ref{FIG_DIAGRAM}(a).
On the lower side of the diagonal,
where the excitatory input becomes more dominant,
all the neurons in the network fire fully synchronized.
On the other hand, in the case
that the inhibitory input dominates the excitatory input,
the clustering state is formed where
the neurons are partially synchronized and each synchronized group
fires asynchronously.
In the diagonal region, the excitatory and inhibitory inputs are balanced, and
the firing pattern of the network is dispersed but not entirely random.

To characterize the structural properties of the complex functional network organized by STDP,
we calculate the clustering coefficient, $C$,
the fraction of connections that actually exist between neighbors of each neurons with respect to all allowable connections,
and the average path length, $L$,
the number of synaptic connections in the shortest path between two neurons averaged over all pairs of neurons.
The phase diagram of the clustering coefficient relative to that of the random network with the same connection probability, $C/C_{rand}$, in Fig.\ref{FIG_DIAGRAM}(c)
is similar to those for the average connection probability and the phase coherence.
In the middle of the diagonal region,
the connection probability is very small with $\left< k/N \right> \approx 0.03$, but
the clustering coefficient for our network is large with $C \approx 0.23$,
whereas the clustering coefficient of the random network, $C_{rand} \approx 0.03$.
In this region, the average path length is $L \approx 3.19$,
while for a random network
$L_{rand} \approx 2.03$.
These results 
show that the functional structure organized by STDP in the case of balanced excitations between the excitatory and the inhibitory coupling has
typical small-world characteristics:
the clustering coefficient of the network is much larger than
that of random network with the same connection probability, $C \gg C_{rand}$,
and the average path length is similar to that of the random network, $L \sim L_{rand}$.

We also find that the degree distributions of the functional structure of the neural network in the case of the balanced input
are scale-free.
Fig.\ref{FIG_DEGREE} shows that
the degree distributions follow a power-law decay
with a cut-off at large k:
$P_{in}(k) \sim k^{-\gamma_{in}}$,
$P_{out}(k) \sim k^{-\gamma_{out}}$ and
$P_{tot}(k) \sim k^{-\gamma_{tot}}$,
where $P_{in}$, $P_{out}$, and $P_{tot}$ are the frequency of nodes with the same number of in-coming, out-going and total synaptic connections independent of directions, respectively.
In the middle of the diagonal region in Fig.\ref{FIG_DIAGRAM}, the scaling exponents are measured to be
$\gamma_{in} \approx 1.7$, $\gamma_{out} \approx 1.4$ and $\gamma_{tot} \approx 1.5$.
The estimated values of $\gamma$ do not depend much on the details of synaptic parameters, $G_{max}$ and $G_{inh}$, around the diagonal region.

\begin{figure}[tb]
\unitlength1cm
\resizebox{6.0cm}{4.0cm}{\includegraphics{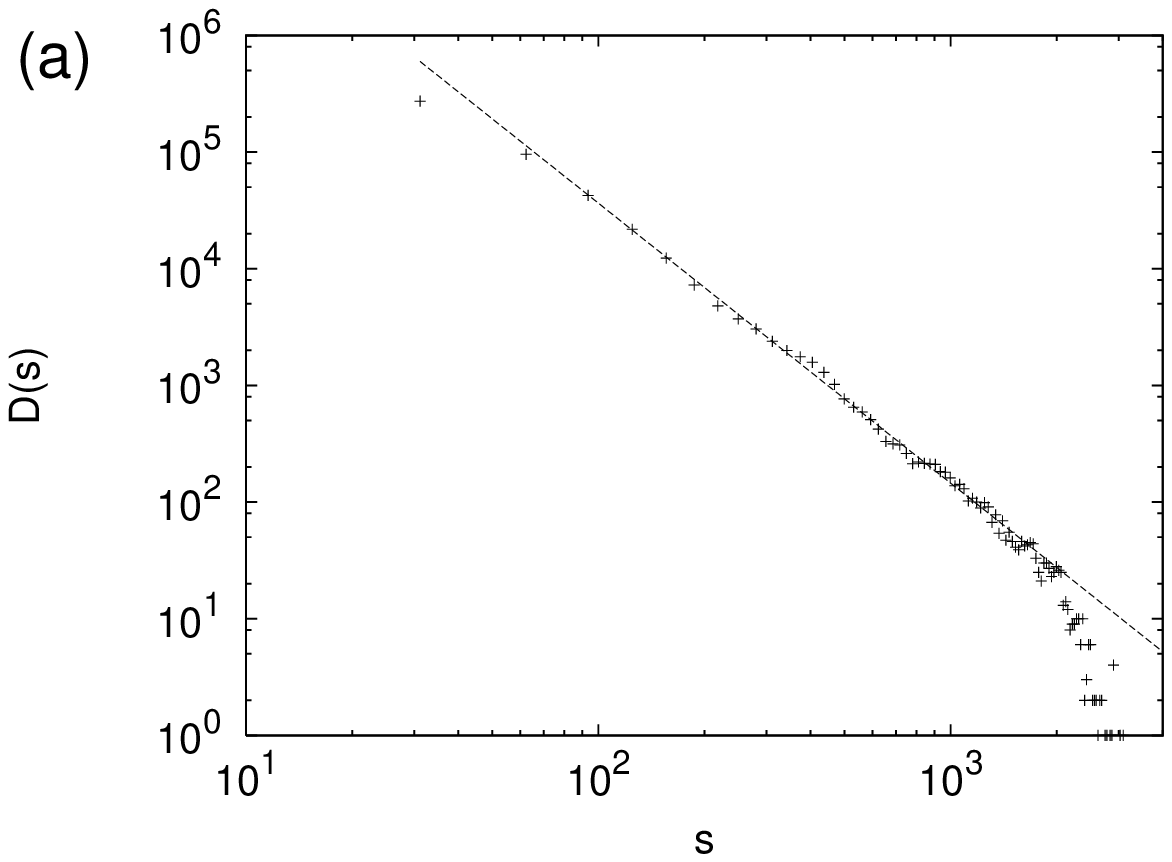}}
\resizebox{6.0cm}{4.0cm}{\includegraphics{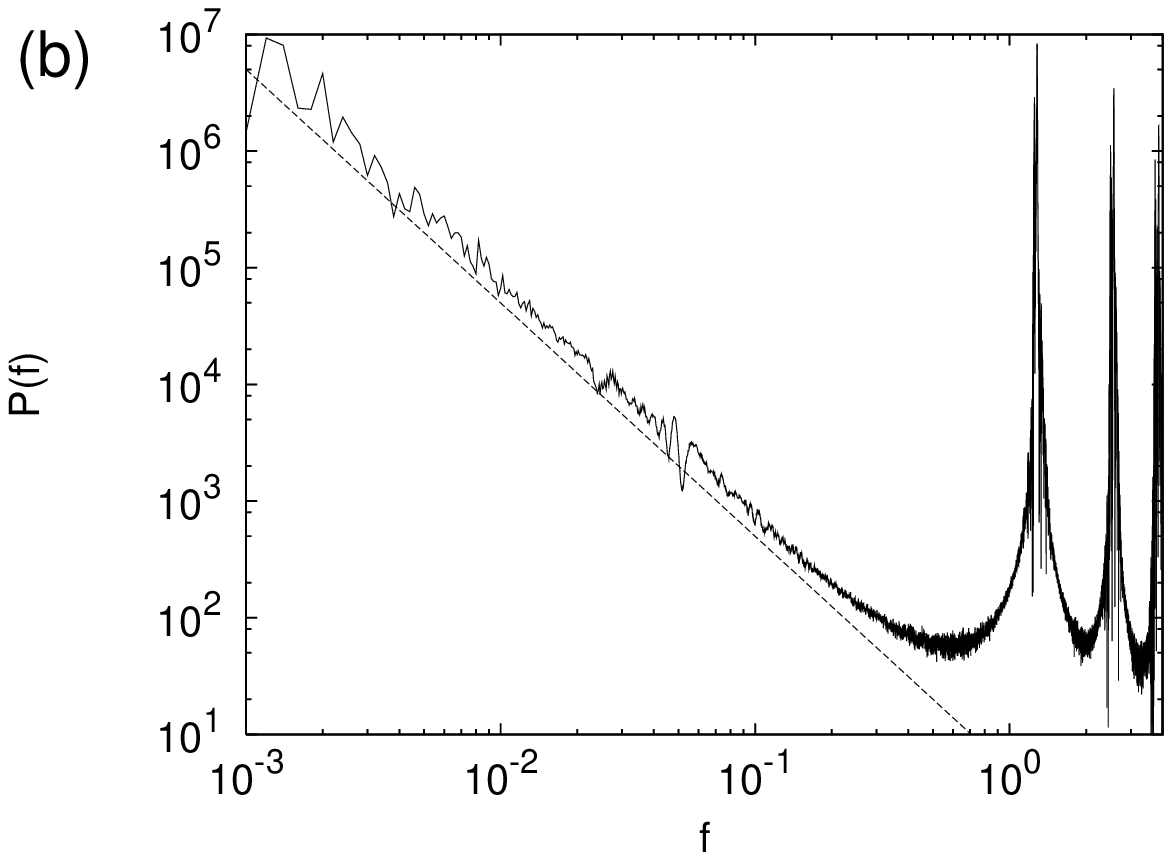}}
\caption {Distribution of the size of the change of the total synaptic coupling strengths (a)
and the power spectrum of the fluctuation of the synaptic strengths (b).
Both follows a power-law decay with exponents, $\gamma_{s} \approx 2.5$ and $\gamma_{f} \approx 2.0$.
The power spectrum has peaks at the natural frequency of the FHN neuron and its harmonics.}
\label{FIG_SOC}
\end{figure}

After a period of relaxation, the average network properties of the functional network remain constant,
but the synaptic coupling strengths continue to fluctuate
as the neurons under the supra-threshold stimulus generate action potentials spontaneously.
Fig.\ref{FIG_SOC} shows that the distribution of the sizes of the change of the total synaptic coupling strength
per unit time, $D(s)$,
and the low-frequency power spectrum of the fluctuation, $P(f)$,
power-law decays, $D(s) \sim s^{-\gamma_{s}}$ and $P(f) \sim 1/f^{\gamma_{f}}$,
with scaling exponents $\gamma_{s} \approx 2.5$ and $\gamma_{f} \approx 2.0$.
These facts suggest that the firing dynamics of the neurons on the small-world scale-free functional network lies in a self-organized critical state,
and the fluctuation is random with no time correlation.
In this critical state,
a slight change in synapses may bring a significant change in dynamical firing patterns,
which can in turn induce a larger change in synapses in an avalanche-like manner
as in the case of the sandpile models \cite{TURCOTTE:1999:REPPROGPHYS}.


Our results show that by STDP the small-world and scale-free functional structure can be spontaneously organized in the neural network
under common external input stimulus, in the form of the self-organized critical state.
The balance between excitation and inhibition in the network dynamics is critical to the formation of the nontrivial network structure.
The experimental studies
using fMRI and MEG in human brain sites also show that
the functional networks in the brain are in fact scale-free small-world networks \cite{EGUILUZ:2003:PREPRINT,STAM:2004:NEUROSCILETT}.
In a small-world network, due to the large clustering and the short average path length,
faster and larger synchronization can be achieved with only a small number of connections \cite{LAGO-FERNANDEZ:2000:PRL,HONG:2002:PRE,MASUDA:2004:BIOLCYBERN},
and the scale-free network is robust against the random failure of nodes \cite{ALBERT:2002:RMP}.
The functional structure by STDP also shows both fast synchronization and high coherence which is dynamically effective and structurally robust.

In the case of conventional Hebbian networks,
if the common external stimulus is given to a part of a neural network,
the synaptic connection strengths between the neurons under the stimulus increase
whereas the other synapses are weakened.
However, out results suggest that even the neurons under common stimulus need not be functionally connected,
but only a small portion of the synapses between neurons can be strengthened
to make the network sparse but small-world and scale-free.
We expect that our work would provide insights on the studies of 
the formation of complex networks
and the developmental process of neural circuits in the brain,
as in the learning and memory models.
We also expect that this neural mechanism could be utilized in controlling the neural network efficiently and enlarging the memory capacity.




\begin{thebibliography}{99}

\bibitem{NEWMAN:2003:SIAMREVIEW}
	M. E. J. Newman, 2003, SIAM Review, {\bf 45} 167.
\bibitem{ALBERT:2002:RMP}
	R. Albert and A.-L. Barab\'{a}si, 2002, Rev. Mod. Phys., {\bf 74} 47.
\bibitem{DOROGOVTSEV:2002:ADVPHY}
	S. N. Dorogovtsev and J.F.F. Mendes, 2002, Adv. Phys., {\bf 51} 1079.
\bibitem{AMARAL:2004:EURPHYSJB}
	L. A. N. Amaral and J.M. Ottino, 2004, Eur. Phys. J. B, {\bf 38} 147.

\bibitem{WHITE:1986:}
	J. G. White, E. Southgate, J. N. Thompson and S. Brenner, 1986,
	Philosophical Transactions of the Royal Society of London, Series B {\bf 314} 1.

\bibitem{MARKRAM:1997:SCIENCE}
	H. Markram, J. K\"{u}bke, M. Frotscher, and B. Sakmann,
	1997, Science {\bf 275} 213.
\bibitem{DEBANNE:1998:JPHYSIOL}
	D. Debanne, B. H. G\"{a}hwiler, and S. M. Thompson,
	1998, J. Physiol. {\bf 507} 237.
\bibitem{BI:1998:JNEUROSCI}
	G.-Q. Bi and M.-M. Poo, 1998, J. Neurosci., {\bf 18} 10464.
\bibitem{BELL:1997:NATURE}
	C. C. Bell, V. Z. Han, Y. Sugawara and K. Grand,
	1997, Nature {\bf 387} 278.

\bibitem{FITZHUGH:1961:BJ}
	R. FitzHugh, 1961, Biophys. J., {\bf 1} 445.

\bibitem{Rossum:2000:JNEUROSCI}
	M. C. W. van Rossum, G. Q. Bi and G. G. Turrigiano, J. Neurosci., {\bf 20} 8812
\bibitem{RUBIN:2001:PRL}
	J. Rubin, D. D. Lee and H. Sompolinsky, 2001, Phys Rev Lett., {\bf 86} 364.
\bibitem{BI:2001:ANNUREVNEUROSCI}
	G.-Q. Bi, M.-M. Poo, 2001, Annu. Rev. Neurosci. {\bf 24} 139.

\bibitem{SONG:2000:NATURENEUROSCI}
	S. Song, K. D. Miller and L. F. Abbott, 2000, Nature Neurosci., {\bf 3} 919.

\bibitem{PIKOVSKY:1997:PHYSICAD}
	A. S. Pikovsky, M. G. Rosenblum, G. V. Osipov and J. Kurths, 1997, Physica D, {\bf 104} 219.

\bibitem{TURCOTTE:1999:REPPROGPHYS}
	Donald L. Turcotte, 1999, Rep. Prog. Phys. {\bf 62} 1377.

\bibitem{EGUILUZ:2003:PREPRINT}
	V. M. Egu\'iluz, D. R. Chialvo, G. Cecchi, M. Baliki, and A. V. Apkarian, 2003, arXiv:cond-mat/0309092
\bibitem{STAM:2004:NEUROSCILETT}
	C.J. Stam, 2004, Neurosci. Lett., {\bf 355} 25.

\bibitem{LAGO-FERNANDEZ:2000:PRL}
	L. F. Lago-Fern\'{a}ndez, R. Huerta, F. Corbacho, and J. A. Sig\"{u}enza,
	2000, Phys. Rev. Lett. {\bf 84} 2758.
\bibitem{HONG:2002:PRE}
	H. Hong, M. Y. Choi, and B. J. Kim, 2002, Phys. Rev. E {\bf 65} 026139.
\bibitem{MASUDA:2004:BIOLCYBERN}
	N. Maduda and K. Aihara, 2004, Biol. Cybern., {\bf 90} 302.

\end{thebibliography}
\end{document}